\documentstyle[12pt,epsfig,cite]{article}

\newcommand{\beq}{\begin{equation}}
\newcommand{\eeq}{\end{equation}}
\newcommand{\bea}{\begin{eqnarray}}
\newcommand{\eea}{\end{eqnarray}}

\newcommand{\gsim}{\lower.7ex\hbox{$
\;\stackrel{\textstyle>}{\sim}\;$}}
\newcommand{\lsim}{\lower.7ex\hbox{$
\;\stackrel{\textstyle<}{\sim}\;$}}

\def\lsim{\mathrel{\rlap{\lower3pt\hbox{\hskip0pt$\sim$}}
    \raise1pt\hbox{$<$}}}         
\def\gsim{\mathrel{\rlap{\lower4pt\hbox{\hskip1pt$\sim$}}
    \raise1pt\hbox{$>$}}}         

\newcommand{\bibit}[1]{\bibitem{#1}}

\newcommand{\aver}[1]{\langle #1\rangle}

\newcommand{\Lam}{\Lambda_{\rm QCD}}

\newcommand{\as}{\alpha_s}
\newcommand{\GeV}{\,\mbox{GeV}}
\newcommand{\MeV}{\,\mbox{MeV}}

\newcommand{\msp}[1]{\mbox{\hspace*{#1mm}~}}

\begin{document}
\thispagestyle{empty}
\vspace*{-10mm}

\begin{flushright}
Bicocca-FT-03-34\\
UND-HEP-03-BIG\hspace*{.08em}08\\
hep-ph/0401063\\
\vspace*{2mm}
\end{flushright}
\vspace*{25mm}

\boldmath
\begin{center}
{\LARGE{\bf
Moments of semileptonic  \boldmath $B$ decay\vspace*{4mm} distributions
in the $1/m_b$ expansion
}}
\vspace*{4mm} 

\end{center}

\unboldmath
\smallskip
\begin{center}
{\Large{Paolo~Gambino}}  \\
\vspace{2mm}
{\sl INFN, Sezione di Torino, 10125 Torino, Italy}  \vspace*{2.5mm}\\
{{\large  and}} \vspace*{3.0mm}
\\
{\Large{Nikolai~Uraltsev$^{*}$
}} \vspace*{2mm} \\
{\sl Department of Physics, University of Notre Dame du Lac,
Notre Dame, IN 46556, USA}\vspace*{-.5mm}\\
{\small {\sf and}}\vspace*{-.5mm} \\
{\sl INFN, Sezione di Milano,  Milano, Italy} 
\vspace*{18mm}

{\bf Abstract}\vspace*{-.9mm}\\
\end{center}
\noindent
We report the OPE-based predictions for a number of lepton energy and
hadronic mass moments in the inclusive semileptonic
$B\!\to\!X_c\,\ell\nu$ decays with a lower cut on the charged lepton
energy. We rely on the direct OPE
approach where no expansion in the charm mass is employed and the
theoretical input is a limited set of underlying OPE parameters
including $m_b$ and $m_c$. A Wilsonian treatment with a `hard'
cutoff is applied using running low-scale masses $m_Q(\mu)$ and
kinetic expectation value $\mu_\pi^2(\mu)$. This leaves for
perturbative corrections only genuinely short-distance effects and
makes them numerically small. Predictions are also given
for the modified hadronic moments of the kinematic variable 
${\cal N}_X^2$ which is a combination of $M_X^2$ and $E_X$. 
Measurement of such 
moments would allow a more reliable 
extraction of higher-order
nonperturbative heavy quark parameters from experiment.

\setcounter{page}{0}

\vfill

~\hspace*{-12.5mm}\hrulefill \hspace*{-1.2mm} \\
\footnotesize{
\hspace*{-5mm}$^*$On leave of absence from 
St.\,Petersburg Nuclear Physics 
Institute, Gatchina, St.\,Petersburg  188300, Russia}
\normalsize

\newpage

\section{Introduction and motivation}

The heavy quark expansion based on the local Operator Product
Expansion (OPE) \cite{vsope,buv,bs} allows to accurately
calculate sufficiently inclusive decay probabilities, incorporating
bound-state and hadronization effects in terms of a limited number of
physical
heavy quark parameters (for details, 
see the review \cite{ioffe}).
Inclusive $B$ decay distributions -- in particular those related to 
$b\!\to\! c\,\ell\nu$ transitions  --
represent a portal to a precise determination of 
these parameters \cite{optical,motion},
and lepton energy moments and moments of hadronic mass and/or
energy are among the most interesting quantities. 
The main ingredients for their computations have been around for 
some time (see 
\cite{buv,prl,koyrakh,volmom,grekap,czarj,gremst,fls95}). Several numerical 
applications can be found in the
literature, for moments with or without cuts on the charged lepton 
energy (see \cite{flcut,DELPHI,gfit} and references therein) often dictated 
by the reality of experimental measurements. 
The present study differs from and complements previous reports in a
number of aspects.

First, we make use of the robust approach advocated in \cite{amst,imprec}; in
particular, we do not invoke an expansion in $1/m_c$ which has plagued
the reliability of  
many earlier applications of the heavy quark
expansion. Moreover, in our approach only the heavy quark
parameters relevant to inclusive decay rates are invoked. This reduces
the number of new objects appearing at order $1/m_Q^3$ from six to two
and eliminates poorly known non-local correlators. An extended set of
experimental moments allows to constrain all relevant parameters
provided model-independent bounds \cite{bounds} are incorporated. 

We also rely on heavy quark parameters which are renormalized {\it a la}
Wilson and depend explicitly on a `hard' normalization scale $\mu$
which we set equal to $1\GeV$.  
Primarily, this refers to the heavy quark masses $m_b(\mu)$ and $m_c(\mu)$ and
to the kinetic expectation value $\mu_\pi^2(\mu)$.
 On the theoretical side,
this is necessary both to meaningfully assign them definite values and
to apply exact heavy quark inequalities. On the practical side this
renders perturbative corrections well-behaved and
moderate in size. Absence of large higher order corrections is
crucial for a meaningful extraction of the nonperturbative parameters.

The direct use of the underlying set of heavy quark
parameters reveals what a particular
moment is actually measuring or constraining, and 
simplifies considerably the task of estimating the theoretical accuracy,
often the subject of
controversial claims \cite{ckm03fpcp}. It is worth reminding
\cite{motion} in this respect that
the inclusive decays cannot depend on non-local correlators often
appearing in other applications of the heavy quark expansion, and
there is no way to constrain them directly studying only inclusive $B$ decays.

The important motivation behind the present study is to make available
a code 
for evaluating the various distributions,
which is   not bound to  specific technical assumptions, 
for instance the use of meson mass relations,  and is flexible enough to 
allow a meaningful investigation of the theoretical uncertainty
using different options (see Section 3).
The detailed analytic expressions will be given in
Ref.~\cite{future}, while here we  limit ourselves with presenting
easy numerical recipes and a  discussion of the main points. 

Finally, we present here predictions for the modified higher hadronic
moments with a charged lepton energy cut, which can be measured in the 
experimental setup of the $B$
factories. They will allow to pinpoint higher-order nonperturbative
expectation values with better accuracy and reliability.

\section{Theoretical setup}

All inclusive semileptonic $B$ decay distributions are described by a 
few  $B$ decay structure functions $w_i(q_0; q^2)$, where $q^\mu$ is the
four-momentum of the lepton pair; for massless
leptons one has \cite{koyrakh}
\bea
\nonumber
\frac{{\rm d}^3 \Gamma}{{\rm d}E_{\ell\,}  {\rm d}q^{2\,} {\rm d}q_0 }
&\msp{-4}= \msp{-4}& \frac{G_F^2 |V_{cb}|^2}{32\pi^4}\,
\vartheta\!\left(q_0\!-\!E_\ell\!-\!\mbox{$\frac{q^2}{4E_\ell}$}\right)
\vartheta(E_\ell) 
\,\vartheta(q^2) \;\times \\ 
& & \msp{20} \left\{
2 q^2 w_1+[4E_\ell (q_0\!-\!E_\ell)\!-\!q^2]w_2 +
2q^2(2E_\ell\!-\!q_0) w_3
\right\} .
\label{12}
\eea
The OPE-based heavy quark expansion yields them 
in terms of (short-distance) quark masses
and of the $B$-meson expectation values of {\it local} heavy quark
operators. Accounting for the latter through $D\!=\!6$, the set of
input parameters for our evaluation of the 
moments includes therefore $m_b$,
$m_c$, $\mu_\pi^2$, $\mu_G^2$, $\tilde\rho_D^3$ and
$\rho_{LS}^3$. Hadronic mass moments kinematically depend also on the
$B$ meson mass $M_B$ for which the experimental value $5.279\GeV$ is
employed. In
practice, it appears in the combination $\bar\Lambda^\prime\!\equiv\!
M_B\!-\!m_b$. As mentioned above, the input heavy quark parameters
depend on the normalization point $\mu$ for which we adopt $1\GeV$.
\footnote{Here we
use, however, the `pole-type' Darwin expectation value $\tilde\rho_D^3
\!\simeq\! \rho_D^3(1\GeV)\!-\!0.1\GeV^3$ instead of the Wilsonian
$\rho_D^3(1\GeV)$, cf.\ Ref.~\cite{imprec}.}

The general structure of the OPE expressions for moments is described
elsewhere \cite{DELPHI,future}. Here we give some technical details of
our calculations. 

Charged lepton energy moments are computed as the ratios
\beq
M_\ell^{(n)}(E_{\rm cut})=\frac{\int_{E_{\rm cut}} E_\ell^n 
\frac{{\rm d}\Gamma}{{\rm d}E_\ell}\, {\rm d}E_\ell}
{\int_{E_{\rm cut}}  
\frac{{\rm d}\Gamma}{{\rm d}E_\ell}\, {\rm d}E_\ell}\;.
\label{102}
\eeq
Both numerator and denominator have been computed through order
$1/m_b^3$ compared to the leading partonic result; perturbative
corrections are included from terms $\alpha_s$ and
$\beta_0\alpha_s^2$, and the ratio has been expanded in both
perturbative and nonperturbative corrections.
Since perturbative corrections in the
Wilsonian scheme are suppressed, an alternative procedure with separate
numerical evaluation of both numerator and denominator would yield a
close result.
Perturbative effects are expressed in terms of 
$\alpha_s^{\overline{\rm MS}}(m_b)$ 
for which we adopt $0.22$ as  central value.
\vspace*{1.5mm}

Hadronic invariant mass squared $M_X^2$ in the OPE appears 
as a special choice
in a one-parameter family of kinematic variables 
\bea
\nonumber
{\cal M}^2(L_\nu)\msp{-4}&=&\msp{-4}
(L_\nu\!+\!m_b v_\nu\!-\!q_\nu)^2=L^2+(m_b v_\nu \!-\!q_\nu)^2+
2 L_\nu(m_b v_\nu\!-\!q_\nu)\,,\\
& &\msp{5} (m_b v_\nu\!-\!q_\nu)^2\equiv m_x^2,\;\;\; (m_b\!-\!q_0)\equiv e_x
\msp{10}
\label{110}
\eea
corresponding to $L_\nu\!=\!\bar\Lambda^\prime v_\nu$, where
$v_\nu\!=\!\frac{P^B_\nu}{M_B}\!=\!(1,{\bf 0})\,$ is the $B$ four-velocity,
and $m_x^2$ and $e_x$ have the meaning of invariant mass squared and
energy in the hadron sector at the quark decay level, 
$e_x\!=\!E_X\!-\!\bar\Lambda^\prime$. The OPE computes
$m_x^2$ and $e_x$ much in the same way as lepton energy moments;
therefore averages of $M_X^2$ and its powers are generally polynomials in 
$(M_B\!-\!m_b)$, with the coefficients computed in the local OPE
\cite{WA}. The latter are various mixed moments $\aver{e_x^k\,
m_x^{2n}}$ derived from Eq.~(\ref{12}).

The variable ${\cal M}^2$ represents an observable for  arbitrary
$L_\nu$. Being a combination of hadronic invariant mass and hadronic
energy (and most generally of spacelike momentum), it can also be viewed as
conventional ``hadronic'' invariant mass square if one considers
not the decay of an isolated $B$ meson, but rather of a compound of $B$
combined with a
non-interacting `spurion' particle having momentum
$L_\nu\!-\!(P^B_\nu\!-\!m_b v_\nu)$. From the OPE perspective, however, the
inclusive probabilities appear as  in the decay of the $b$ quark, while the
momentum associated with the light cloud looks like a spurion. The
native object for the OPE is therefore  ${\cal M}^2(L_\nu)$ with
vanishing $L_\nu$, rather than $L_\nu\!=\!\bar\Lambda^\prime v_\nu$
which yields $M_X^2$.  

It turns out that for higher hadronic moments the generalized moments
with $L_\nu \!\ll\! \bar\Lambda^\prime$ are advantageous, they are better
controlled theoretically and more directly sensitive to 
higher-dimension expectation values.\footnote{This roots to lower 
infrared sensitivity of the modified  moments 
compared to those of $E_X$ which, in higher orders, are dominated
by maximal-$q^2$ kinematics where the $c$ quark is nearly at rest. The 
combinatorial factors for the terms $\bar\Lambda^\prime e_x$ in the
conventional $M_X^2$ moments work in the same direction, 
being additionally enhanced by 
large value of $\bar\Lambda^\prime$.}
To utilize this advantage we consider,
along with $M_X^2$ the modified moments, i.e.\ those of 
\beq
{\cal N}_X^2=M_X^2-2\tilde\Lambda E_X+\tilde\Lambda^2
\label{114}
\eeq
with $\tilde\Lambda\!=\!0.65\GeV$,  close to the anticipated
value of $\bar\Lambda^\prime$. In this case ${\cal N}_X^2\!-\!m_c^2$ approximates
the {\sf quark} virtuality, for which higher moments with respect to
average are intrinsically related to higher-dimension expectation
values. (The last constant term in Eq.~(\ref{114}) does not affect
such moments.) In our approach computing these modified moments does not
require a new analysis -- they are given by simply replacing $M_B$ by
$M_B\!-\!\tilde\Lambda$. 

The higher moments of the decay distributions -- in particular 
those of the hadronic invariant mass  -- are more informative 
when considered with respect to the average, say
$\aver{(M_X^2\!-\!\aver{M_X^2})^2}$,
$\aver{(M_X^2\!-\!\aver{M_X^2})^3}$, 
or similar moments for ${\cal N}_X^2$. These moments are  the focus of our
study. Since this may complicate to some extent the
experimental error analysis,\footnote{We thank
O.~Buchmueller for discussing this point and alternative options.} we
also present similar numerical results for the moments evaluated with
respect to a fixed hadronic mass, for which we take $4\GeV^2$ in the case
of $M_X^2$-moments and $1.35\GeV^2$ for ${\cal N}_X^2$ ones. 

Power corrections in moments with lepton
cut are obtain directly integrating the published 
heavy quark structure functions. The emerging analytic expressions 
are not too complicated, but lengthy and consist of many terms,
especially for higher hadronic moments. We will present them in a
dedicated publication \cite{future}. Perturbative corrections are
cumbersome and require numerical integrations. At the same time, in our
approach utilizing a `hard' separation between short- and
long-distance effects they are numerically quite small. 
(For instance, it follows from the tables in the Appendix that 
including perturbative corrections in $\aver{M_X^2}$ has the same impact
as decreasing $m_b$ by $15\MeV$, or as decreasing $\mu_\pi^2$ by
$0.02\GeV^2$ in $\aver{(M_X^2\!-\!\aver{M_X^2})^2}$.)
Therefore we evaluate hadronic moments to the first order in $\alpha_s$
using the value $\alpha_s\!=\!0.3$ (average gluon virtuality in $B$
decays is lower than $m_b$). Moreover, in the present paper we
evaluated perturbative shifts in the hadronic moments neglecting the
cut on the lepton energy. This seems to be a legitimate approximation
since at $E_{\rm cut}\!=\!0$ they are small and do not exceed the
expected accuracy, which is limited by other neglected effects. This
element will be improved in Ref.~\cite{future}.

Unlike  the case of lepton moments, for power-suppressed terms we do not
use the expanded form of the ratios that form the moments, nor drop 
any power-like terms generated by $\mu_\pi^2$, $\mu_G^2$, $\rho_D^3$
and  $\rho_{LS}^3$  wherever they appear. In particular, this implies
that we would not include $M_B\!-\!m_b$ (or its analogue
$M_B\!-\!m_b\!-\!\tilde\Lambda$ for ${\cal N}_X^2$-moments) into
counting powers of $\Lam$. This is natural since that entry is 
external to the OPE
for inclusive probabilities and it can take arbitrary values, both much
larger and much smaller than $\Lam$, as illustrated by the modified
hadronic moments. Similarly, in the perturbative corrections terms
like $\sim \!\alpha_s\mu_\pi^2$,  $\;\alpha_s\mu_G^2$ etc.\ are not retained,
but those $\sim \!\alpha_s \bar\Lambda^k$ without nonperturbative expectation
values are legitimately kept for arbitrary power $k$.\footnote{Terms 
\,$O(\alpha_s \bar\Lambda'^k)$\, with $k\!>\!0$ 
have not been included in the third hadronic moment.}

In this note we present our numerical results in a simplified
form, using the following reference  values of the parameters
$$
m_b\!=\!4.6\GeV,\;\;\; m_c\!=\!1.2\GeV,\;\;\; \mu_\pi^2\!=\!0.4\GeV^2,\;\;\;
\mu_G^2\!=\!0.35\GeV^2,\;
$$
\vspace*{-7mm}
\beq
\tilde\rho_D^3\!=\!0.1\GeV^3,\; \;\;\;
\rho_{LS}^3\!=\!-0.15\GeV^3,
\label{120}
\eeq
and providing the coefficients for the linear extrapolation in these
values from this base point. Such linearized extrapolations appear
sufficiently accurate for reasonable values of the parameters. 
The tables summarizing our results are given in the Appendix.
Yet this is no more than the simplest compact way of
communicating our results (for instance, we drew plots using the complete
expressions rather than interpolations). The numerical evaluations 
for other central
values, more accurate interpolating tables, or compact Mathematica or
FORTRAN programs evaluating them are available upon request, and will
be provided with Ref.~\cite{future}.

Measured in experiment are also non-integer moments of $M_X^2$,
most notably  $\aver{M_X}$ and $\aver{M_X^3}$
\cite{babarpr}.
They do  not arise naturally  in the $1/m_b$ expansion,
as illustrated by the limit $m_c\!\to\! 0$ which is 
analogous to the decay $B\!\to\! X_s+\gamma$; fractional photon energy 
moments are not given there 
by the expectation values of local heavy quark operators. For $B\!\to\!
X_c\,\ell\nu$,  the OPE would involve an expansion in
$1/m_c$, as can also be seen from
\beq
\aver{M_X^\nu}= \left(\aver{M_X^2}\right)^{\frac{\nu}{2}} \left[1 + 
\sum_{k=2}^\infty 
\raisebox{-2pt}{\mbox{{\large$C$}}}_{\!_{\frac{\nu}{2}}}^{^{\,k}} 
\,\frac{\aver{(M_X^2-\aver{M_X^2})^k}}{\aver{M_X^2}^k}
\right];
\label{180}
\eeq
for integer moments with $\nu\!=\!2n$ the sum contains only terms through
$k\!=\!n$ and stops before $\aver{M_X^2}$ enters the denominator.

Having computed three first (integer) hadronic moments, we truncated
the sum in Eq.~(\ref{180}) after $k\!=\!3$. Although incomplete, 
at the actual value of the charm mass
this truncated expansion appears a 
sufficiently good numerical approximation for $\nu\!=\!1$ and 
$\nu\!=\!3$; 
the omitted terms seem significantly below the actual
theoretical accuracy in evaluating the integer moments involved.

In fact, $M_X^2$ contains the dominant term $2\bar\Lambda^\prime e_x$
as well as the perturbative bremsstrah\-lung contribution, and both
effects can be computed explicitly for arbitrary  $\nu$ without truncating
the series in Eq.~(\ref{180}). The numerical impact of this
resummation turns out totally negligible numerically. Moreover, while the
term  $2\bar\Lambda^\prime e_x$ dominates  $\aver{M_X^2}$, it is 
subdominant in the higher moments with respect to  $\aver{M_X^2}$.

\thispagestyle{plain}
\vspace*{.3mm}
\begin{figure}[hhh]
\mbox{\epsfig{file=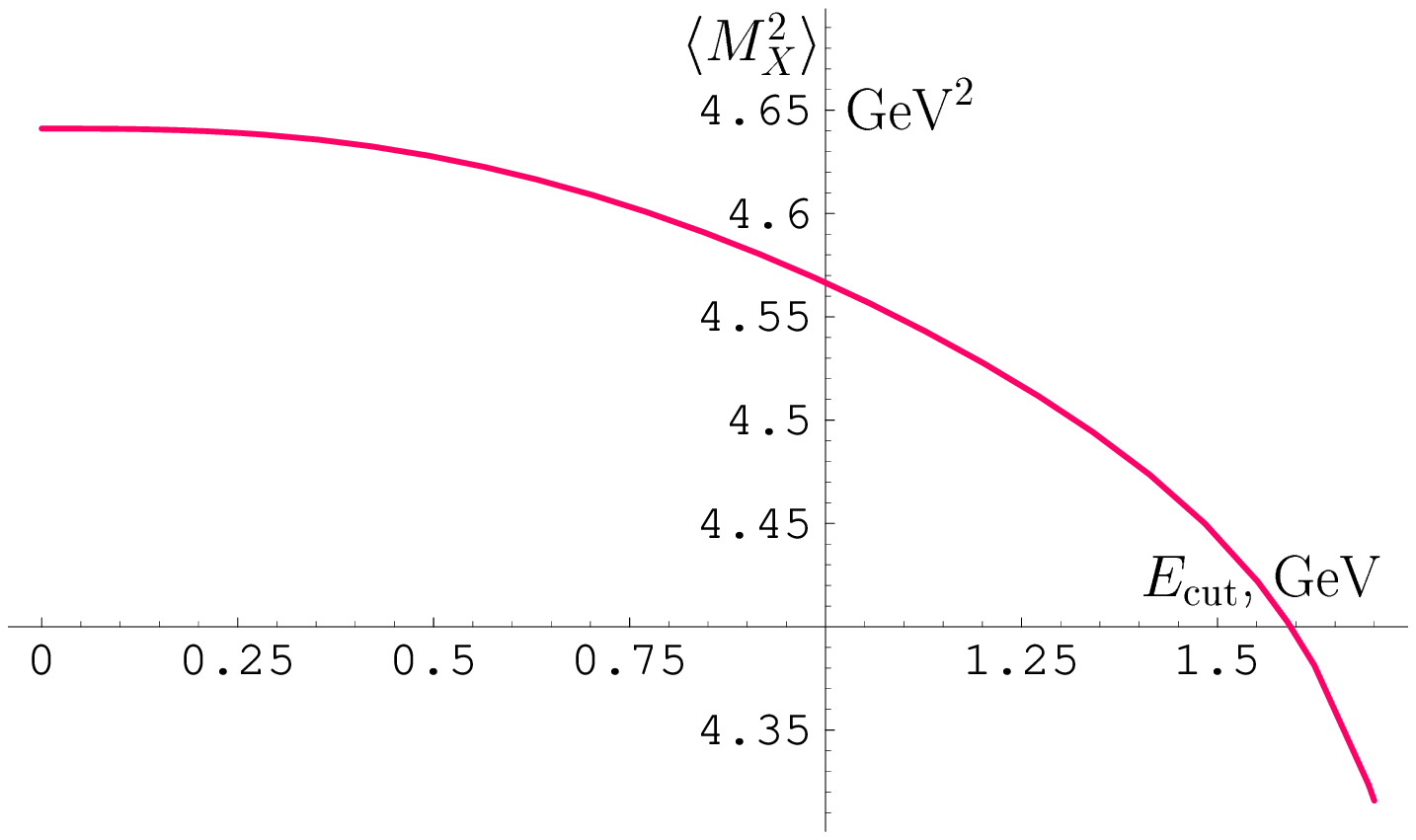,width=7.2cm}}
\hfill
\mbox{\epsfig{file=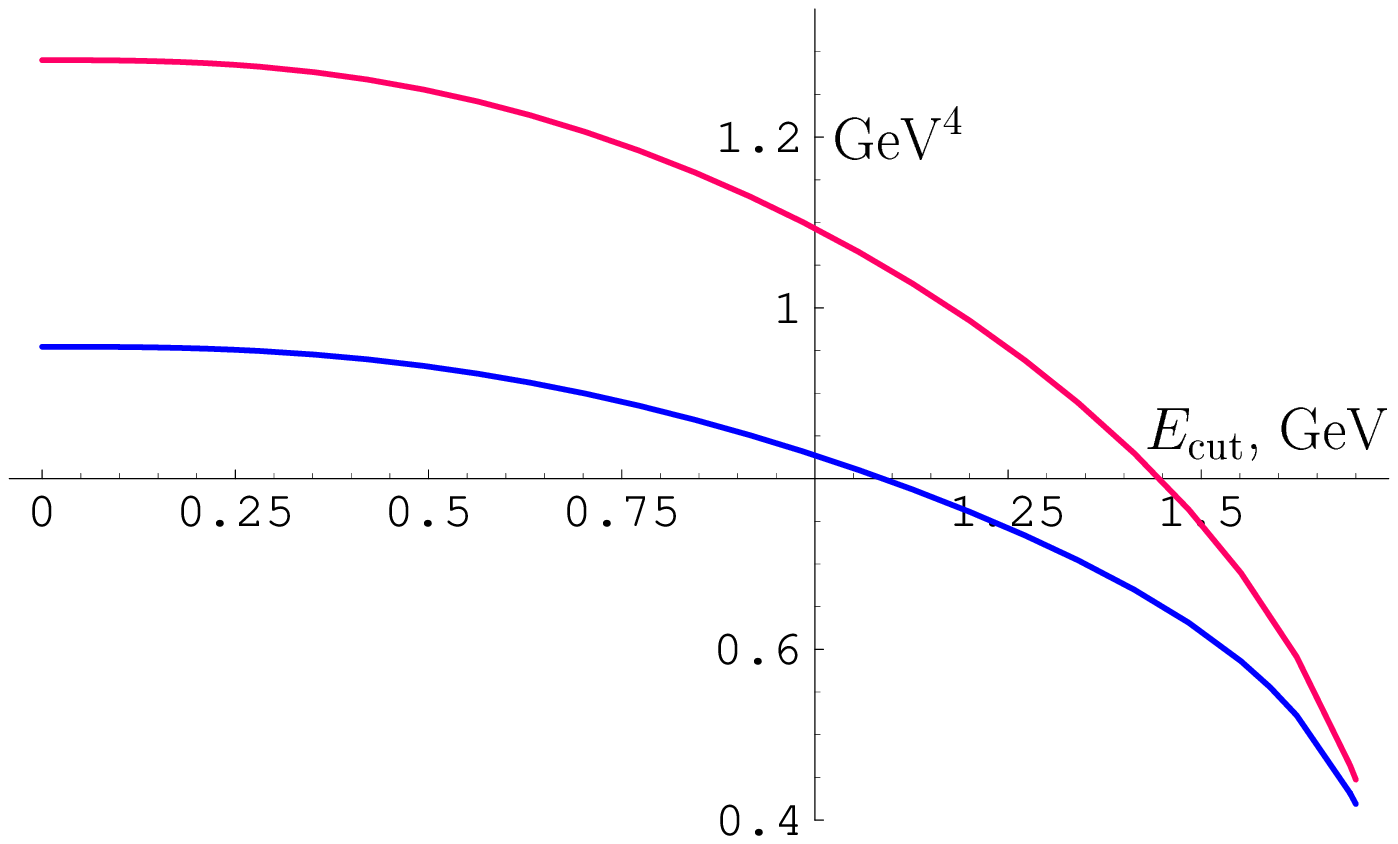,width=7.2cm}}\vspace*{-4.5mm}\\
\begin{minipage}[t]{7.2cm}
\caption{ \small
Average hadronic invariant mass squared $\aver{M_X^2}$
at different lepton energy cuts, for the 
heavy quark parameters of Eq.~(\ref{120}).
}
\end{minipage} \hfill
\begin{minipage}[t]{7.2cm}
\caption{ \small
Second invariant mass moment $\aver{[M_X^2\!-\!\aver{M_X^2}]^2}$ (red) and
second modified hadronic moment  
$\aver{[{\cal N}_X^2\!-\!\aver{{\cal N}_X^2}]^2}$ (blue), in the same
setting.
}
\end{minipage} 
\end{figure}

For illustration we give a few plots showing dependence of the hadronic mass
moments on the lepton energy cut, at the central values of heavy quark
parameters. Fig.~1 depicts $\aver{M_X^2}$, Fig.~2 addresses  
$\aver{(M_X^2-\aver{M_X^2})^2}$ and
$\aver{({\cal N}_X^2-\aver{{\cal N}_X^2})^2}$. Non-integer $M_X^2$ 
moments are shown in
Fig.~3 where we actually plot the corresponding powers of the moments
having dimension of mass,  for $\nu=1,\,2,\,3,\,4,\,5$
and $6$. It shows that the full moments are by far dominated by the
average invariant mass, with relatively small differences.

\thispagestyle{plain}
\begin{figure}[hhh]\vspace*{-3.4mm}
\begin{center}
\mbox{\epsfig{file=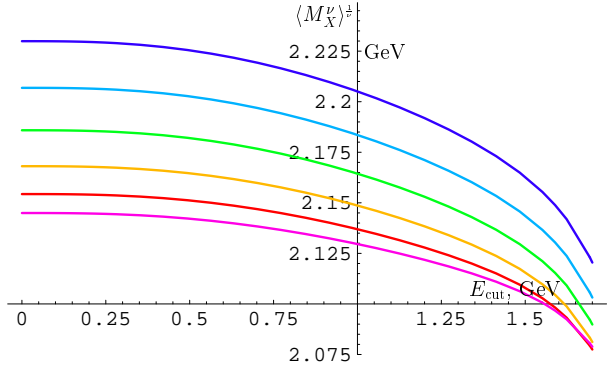,width=8cm}}
\end{center}
\vspace*{-6.0mm}
\caption{ \small
Different hadronic mass moments $\aver{M_X^\nu}^{\frac{1}{\nu}}$ for
$\nu\!=\!1$ to $6$ (from lowest to highest), 
vs.\ lepton energy cuts; heavy quark parameters as in Eq.~(\ref{120}).
}
\end{figure}

As the  semileptonic decay rate  is well measured only above
certain energy $E_\ell$,  the OPE predictions for the decay 
fraction 
\beq
R^*(E_{\rm cut}) = \frac{\int_{E_{\rm cut}} 
\frac{{\rm d}\Gamma}{{\rm d}E_\ell}\, {\rm d}E_\ell}
{\int_{0}  
\frac{{\rm d}\Gamma}{{\rm d}E_\ell}\, {\rm d}E_\ell}
\label{190}
\eeq
 are helpful to reconstruct the overall
$b\!\to\! c\,\ell\nu$ rate. Since the fraction of events cut out is
small and they belong to the domain of low $E_\ell$ which is 
theoretically most robust, 
the predictions for $R^*$ are expected to be quite reliable.

\section{Discussion}

In the present paper we provide  numerical expressions for the
local OPE predictions of the charged lepton energy and hadronic mass
moments with a lower cut on lepton energy. Two major aspects of our
analysis have already been emphasized. First,
no $1/m_c$ expansion is involved at
any stage, and  the $c$ quark mass can more or less be arbitrary, 
large or small 
within the same formalism.
The second aspect is that we rely on well-defined running quark masses and
$\mu_\pi^2$ normalized at the scale $1\GeV$. (The scheme is often
referred to as `kinetic', however this name is elucidating 
only when applied to quark masses.) The advantages of this approach are
well known, and can be readily seen from the applications presented
here. 

A quick comparison \cite{misuse} with  recent data
\cite{newdelphi,newbabar,newcleo} 
from DELPHI, BaBar and CLEO suggests
agreement between this implementation of the OPE and experiment, with
preferred values of the heavy quark parameters in the range
expected theoretically. (Our estimates show that the expectations for
two moments of the photon energy spectrum appear to fit the values
reported by CLEO \cite{cleobsg} once the `exponential' effects
discussed in Ref.~\cite{misuse} are adjusted for.) The dependence of the
first and second hadronic mass squared moments on the lepton 
energy cut seems to be in a qualitative agreement with the
preliminary data reported by BaBar and CLEO 
\cite{newbabar,newcleo}. The dedicated data analysis 
including possible fits to the predictions  should, in our
opinion be left for experiment, and we did not attempt this.

Once the heavy quark parameters are extracted from data, one can
readily determine, for example, $|V_{cb}|$ from the measured
semileptonic $b\!\to\!c$ decay width \cite{imprec}:
{\small
\begin{eqnarray}
\nonumber
\frac{|V_{cb}|}{0.0417} \msp{-3}&=&\msp{-3} 
\mbox{{\large$
\left(\frac{{\rm Br}_{\rm sl}(B)}{0.105}\right)^{\!\frac{1}{2}} 
\left(\frac{1.55{\rm ps}}{\tau_B}\right)^{\frac{1}{2}}$}}
\left(1\!-\!\mbox{{\small$4.8\,[{\rm Br}(B\!\to\!X_u\,\ell\nu)-0.0018]$}}
\right) \cdot 
[1+0.30\,(\alpha_s(m_b)\!-\!0.22)]\,  \nonumber 
\\ 
& & \msp{22}
\times \,\left[1 -0.66 \left ( m_b(1\GeV) \!-\!4.6\GeV \right )
+0.39 \left ( m_c(1\GeV) \!-\!1.15\GeV \right )\qquad \right. \nonumber \\
& & \msp{26}+0.013 \left ( \mu_{\pi}^2 \!-\!0.4\GeV^2 \right )
+0.09 \left ( \rho_D^3(1\GeV) \!-\!0.2\GeV^3 \right ) \nonumber \\
& & \msp{26}\left. +\,0.05 \left ( \mu_G^2\!-\!0.35\GeV^2 \right ) 
-0.01 \left ( \rho_{LS}^3 \!+\!0.15\GeV^3 \right ) \right ]\;.
\label{vcb2}
\end{eqnarray}
}
\hspace*{-6.1pt}It should be recalled, however,
 that $\rho_D^3(1\GeV)$ appearing in the above equation
is related to $\tilde\rho_D^3$ 
employed in the present study by 
$\tilde\rho_D^3\simeq \rho_D^3(1\GeV)\!-\!0.1\GeV^3$.

We do not intend to address here the question of theory
accuracy for the moments to an extent commensurate to the
analysis \cite{imprec} of total $\Gamma_{\rm sl}(b\!\to\! c)$. 
Nevertheless, the numerical results presented in the Appendix
allow a straightforward, if simplified, estimate of possible  
theory inaccuracies. 

Since no perturbative corrections to the Wilson coefficients of
nonperturbative operators have been computed so far, one can assume a
related $\sim \!20\%$ uncertainty in the  contributions due to
$\mu_\pi^2$ and $\mu_G^2$ and a $\sim \!30\%$ uncertainty 
in those due to $\rho_D^3$ and $\rho_{LS}^3$.
As  perturbative corrections in hadronic moments are presently implemented only
at first order in $\alpha_s$,
the associated uncertainty 
can be assessed by varying $\alpha_s$ in a 
reasonable range; since the actual {\it short-distance} corrections
selected by our Wilsonian treatment are  moderate, we may conservatively vary
the effective $\alpha_s$ 
between $0.2$ and $0.45$. To
safeguard against possible accidental cancellations in perturbative
corrections, one may assume an additional minimal uncertainty in $m_{b,c}$
of about $20\MeV$. A similar `minimal uncertainty' in $\mu_\pi^2$ is
probably around $0.02\GeV^2$. 

The results we provide in the Appendix can be improved 
in several ways. The relevance of the improvements is essentially
determined by the state of experiment. A more complete calculation of
perturbative corrections in hadronic moments fully incorporating cuts
in the lepton energy is the first on the list. All-order BLM resummation
can be implemented in both lepton and hadron moments. There are ways
to partially incorporate non-BLM second-order corrections without
performing extensive new calculations. All these improvements in the
perturbative corrections are not expected to essentially modify the
final result in the Wilsonian scheme (see, e.g.\ the dedicated analysis of
$\Gamma_{\rm sl}$ in Ref.~\cite{imprec}). Yet having them incorporated
would add confidence in the present estimates and may allow to 
reduce the theoretical uncertainty.

A probe of the significance of second- and higher-order {\it non-BLM}
corrections is already available by varying the normalization scale
$\mu$ used for quark masses and $\mu_\pi^2$, $\rho_D^3$, while
simultaneously running their values according to \cite{dipole}
\bea
\nonumber
\frac{{\rm d}m_Q(\mu)}{{\rm d}\mu} &\msp{-4}=\msp{-4}& \!
\mbox{$
 -\frac{16}{9}\frac{\alpha_s(M)}{\pi}\, \left(1+\frac{\as}{\pi} 
\left[\frac{\beta_0}{2}
\left(\ln{\frac{M}{2\mu}} + \frac{5}{3}\right)-
3\left(\frac{\pi^2}{6}-\frac{13}{12}\right)\right]\right)$}\\
\nonumber
&& \mbox{$ 
-\frac{4}{3} \frac{\alpha_s(M)}{\pi} \frac{\mu}{m_Q}\left(1+\frac{\as}{\pi} 
\left[\frac{\beta_0}{2}
\left(\ln{\frac{M}{2\mu}} + \frac{5}{3}\right)-
3\left(\frac{\pi^2}{6}-\frac{13}{12}\right)\right]\right)$} 
\; , \\
\nonumber
\frac{{\rm d}\mu_\pi^2(\mu)}{{\rm d}\mu}
&\msp{-4}=\msp{-4}& \msp{2}\mbox{$ \frac{8}{3} 
\frac{\alpha_s(M)}{\pi} \mu \,\left(1+\frac{\as}{\pi} 
\left[\frac{\beta_0}{2}
\left(\ln{\frac{M}{2\mu}} + \frac{5}{3}\right)-
3\left(\frac{\pi^2}{6}-\frac{13}{12}\right)\right]\right)$}\; , \\
\frac{{\rm d}\rho_D^3(\mu)}{{\rm d}\mu}
&\msp{-4}=\msp{-4}& \msp{2} 4 
\mbox{$
\frac{\alpha_s(M)}{\pi} \mu^2 \left(1+\frac{\as}{\pi} 
\left[\frac{\beta_0}{2}
\left(\ln{\frac{M}{2\mu}} + \frac{5}{3}\right)-
3\left(\frac{\pi^2}{6}-\frac{13}{12}\right)\right]\right)$}
\;,
\label{306}
\eea
where $M$ denotes the scale used for normalizing $\alpha_s$.
In fact, in the future  we will also explicitly adopt  
the Wilsonian $\rho_D^3(\mu)$ instead of  $\tilde\rho_D^3$ currently used. 
Even the uncertainty obtained in this way may,
however, not fully represent the uncalculated non-BLM corrections.
A further option is using the $\overline{\rm MS}$ scheme for the charm
mass, $\bar{m_c}(m_c)$. With this choice, one  
could  vary $\mu$ in a wider range to probe 
more quantitatively the actual hardness of a particular moment at a
given $E_\ell$ cut, and a more direct comparison with 
$m_c$ extracted from different physical processes \cite{cernckmrep} 
would be possible. Since no
constraints is imposed on $m_b\!-\!m_c$ through conventional 
mass relations like
\beq
m_b\!-\!m_c = \frac{M_B+3M_{B^*}}{4}-\frac{M_D+3M_{D^*}}{4} +
\mu_\pi^2\left(\frac{1}{2m_c}\!-\!\frac{1}{2m_b}\right) + {\cal O}
\left(\frac{1}{m_Q^2}\right)+ ...
\label{308}
\eeq
the normalization point -- or even the scheme itself -- can be
different for $m_c$ and $m_b$. 

It should be emphasized that we have quoted in the Appendix
what literally emerges from our OPE expressions, 
regardless of whether the cut on $E_\ell$ is mild or severe.
However, if the cut is high, the effective
hardness of the inclusive process degrades and the accuracy
deteriorates. As has been argued \cite{ckm03fpcp,misuse}, the truncated
expressions of practical OPE may not fully reflect this, while the
so-called `exponential' effects may become significant when $E_{\rm
cut}$ exceeds $1.3$ to $1.5\GeV$. The same reservation applies to the
above simplified way to estimate theoretical accuracy -- it largely leaves
out this aspect. The additional cut-related theory errors are possibly
insignificant below $1.35\GeV$; they may not dominate even at 
$E_{\rm cut}=1.5\GeV$ -- yet this cannot be confidently derived from
theory {\it a priori.}\, The safest way to tackle them is to use as
low $E_{\rm cut}$  as possible, not larger than $1.2\GeV$. 

A potential source of additional nonperturbative corrections are the 
so-called `intrinsic charm', or IC contributions in the OPE, associated
with non-vanishing expectation values of four-quark operators with
charm fields $\bar{b}\Gamma c\, \bar{c} \Gamma b$. No allowance is
presently reserved for them. It has been noted in Ref.~\cite{imprec}
that their effect can mimic to some extent the contribution of the
Darwin operator. We think therefore that at the moment theoretical
constraints on $\rho_D^3$ should be used cautiously in the context of
fits based on the OPE without possible IC contributions. 
\vspace{2mm}

There are some conclusions we draw from our  analysis. The
moments under consideration seem to reliably (over)constrain a
certain combination of the two heavy quark masses,  approximately
$m_b\!-\!0.6m_c$; the sensitivity to the individual masses is not  high
and they are subject to larger theoretical uncertainties. 

The moments are only weakly sensitive to the spin-orbit expectation value
$\rho_{LS}^3$, so it cannot be extracted from the data.
On the other hand, $\rho_{LS}^3$ is reasonably constrained  by a number
of exact heavy quark sum rules which place model-independent
bounds. 
Once they are taken into account, the associated uncertainty in the moments
appears by far subdominant. 
Therefore, $\rho_{LS}^3$ does not have to be included in the fit
and we suggest to use a fixed value 
$\rho_{LS}^3\!=\!-0.15\GeV^3$, and to vary it within $\pm
0.1\GeV^3$ to conservatively estimate the related uncertainty. 
This assumption would be further
reinforced if the fits of data prefer values of the
primary heavy quark parameters in  the expected ranges
\bea
\nonumber
&& m_b(1\GeV)=4.60\pm 0.06 \GeV, \qquad \mu_\pi^2(1\GeV)=0.45\pm
0.1\GeV^2, \\
&& 0 < \tilde\rho_D^3<0.15\GeV^3 \;\;\;\mbox{ or }\;\;\;
0.1\GeV^3 < \rho_D^3(1\GeV) < 0.25\GeV^3\;,
\label{320}
\eea
see Refs.~\cite{ioffe,amst,four}; an updated review of different determinations
of beauty and charm quark masses can be found in Ref.~\cite{cernckmrep}.
Similarly, since $\mu_G^2$ is accurately known, it should 
be set to $0.35\GeV^2$ 
\cite{chrom} and  varied  within $\pm 20\%$ to allow for the perturbative
uncertainty in its Wilson coefficient.
\vspace{2mm}

Obtaining informative model-independent
constraints on higher-dimension heavy quark parameters 
requires measurement of 
higher hadronic moments, at least the second and desirably
the third one. 
Since these higher moments -- when considered  with respect to the average --
depend strongly on higher-dimension expectation values, even 
a rough measurement of 
the variance and of the asymmetry parameter
would yield precious information. One should be warned, however, 
that the theoretical accuracy one can
realistically achieve for the higher moments of $M_X^2$ 
is limited, especially in the case of the third moment.
On the other hand,  the modified higher hadronic
moments $\aver{({\cal N}_X^2 \!-\!\aver{{\cal N}_X^2})^2}$, 
$\aver{({\cal N}_X^2 \!-\!\aver{{\cal N}_X^2})^3}$ are better in
this respect and more suitable to this purpose. We encourage
experiments to pay closer attention to such combinations of $M_X^2$
and $E_X$ moments. 
\vspace{2mm}

The actual assessment of the theoretical error in the calculation of  the
moments is a very subtle issue. 
We cannot do without mentioning a few important aspects. 

First, there are 
strong correlations among theoretical
uncertainties. After all, everything inclusive we compute is expressed
in terms of only three structure functions which have physical properties
like positivity, regardless of any dynamics. 
Because of these correlations, one should distinguish between the 
overall consistency of the fits to the various moments in the context
of our formalism, and the concrete prediction for a certain moment.
There are common uncertainties, like a systematic bias in $m_b$ or 
$\mu_G^2$, that would simply shift the fitted value of $m_b$ or 
$\mu_G^2$, but would not degrade the quality of
the global fit, nor would they alter significantly the 
$E_{\rm cut}$-dependence \cite{ckm03fpcp}. At the same time, these uncertainties
could still noticeably affect the numerical value of a particular
moment. 

On the other hand, there are uncertainties that affect each
moment in a different way, for instance, 
unknown perturbative corrections to the Wilson coefficients of
the nonperturbative operators, 
which are in principle different for every
moment, and also depend on  $E_{\rm cut}$. 
Therefore, simply varying the values of the heavy quark parameters
uniformly in all observables may not represent realistically the
uncertainty of the theoretical expressions, and an intermediate
procedure may be required.

Having mentioned these complications of the
theory error analysis, we  
would like to make a 
few suggestions that  can be inferred from our study. 
For sufficiently low cut on $E_\ell$ a reasonable starting point  is
to estimate the theoretical accuracy in
the moments just varying the values of the heavy quark parameters they
depend upon in the ranges we have mentioned above Eq.~(\ref{306}). 
For higher moments, however, this should be applied to the moments with
respect to average; the ordinary moments (around zero) would in this
way exhibit strong theory error correlations. 
Clearly, allowance should be made
for {\it additional} uncertainty once the cut is raised beyond $1.35\GeV$.
\vspace*{2mm}

In summary, we have seen that 
consistency checks between theory and emerging
data should  rely on robust theoretical elements, with additional
assumptions reduced to minimum. Once agreement with theory is
confirmed, and the domain and degree of applicability is verified
experimentally (e.g., the safe interval for lepton cut), one can and
should implement into the  fit of data all the model-independent
relations and bounds following from heavy quark sum rules. These are
exact relations derived in QCD, and they are indispensable 
for a precision determination of the heavy quark parameters and for a 
stringent test of our theoretical tools. 
\vspace*{4mm}

\noindent
{\bf Acknowledgments:} We are indebted to many experimental colleagues from
BaBar, in particular to Oliver Buchmueller and Urs Langenegger for
important discussions which to a large extent initiated the present
and further ongoing studies, and for important physics comments. The early
stages of this investigation ascend to the analysis of LEP data, 
and we are grateful to
DELPHI members, especially to M.~Battaglia, M.~Calvi and
P.~Roudeau for collaboration. N.U.\ is thankful to I.~Bigi for many
important discussions and suggestions. P.G.\ thanks  the Theory
Division of CERN where  this work has been started. 
This work was supported in part by the NSF under grant number
PHY-0087419 and  by a Marie Curie Fellowship, 
contract No.~HPMF-CT-2000-01048.


\renewcommand{\theequation}{A.\arabic{equation}}
\renewcommand{\thesection}{}
\setcounter{equation}{0}
\setcounter{section}{0}
\section{Appendix}
\renewcommand{\thesubsection}{A.\arabic{subsection}}

The following tables give our numerical estimates for various moments. The
general form for a generic moment ${\cal M}$ in question will be
{\small\bea
\nonumber
{\cal M}(m_b,m_c,\mu_\pi^2,\mu_G^2, \tilde\rho_D^3,
\rho_{LS}^3;\alpha_s) \msp{-4}&=&\msp{-4}
V + B\,(m_b\!-\!4.6\GeV)+C\,(m_c\!-\!1.2\GeV) \\
&&\msp{-.9}+ P\,(\mu_\pi^2\!-\!0.4\GeV^2)
+D\,(\tilde\rho_D^3\!-\!0.1\GeV^3) \qquad 
\label{310}
\\
\nonumber
&&\msp{-.9}+  G\,(\mu_G^2\!-\!0.35\GeV^2)
+L\,(\rho_{LS}^3\!+\!0.15\GeV^3)+S\,(\alpha_s\!-\!0.22)\;;
\eea}
\vspace*{-5.5mm}

\noindent
$V$ represent the reference values obtained for the heavy quark parameters 
in Eq.~(\ref{120}). They have dimension of the moment ${\cal M}$ itself
and the quoted number is in $\GeV$ to the corresponding power. The values
of all the coefficients $B$ to $S$ are likewise in the proper power of
$\GeV$ (the same power for $S$, one power less for $B$ and $C$, two
powers less for $P$ and $G$, three powers less for $D$ and
$L$). Values of $E_{\rm cut}$ are shown in $\GeV$ as well.

All the moments are given without cut on $E_\ell$ (i.e.\ $\,E_{\rm cut}\!=\!0$)
and for $E_{\rm cut}\!=\!0.6\GeV$, $E_{\rm cut}\!=\!0.9\GeV$, 
$E_{\rm cut}\!=\!1.2\GeV$ and $E_{\rm cut}\!=\!1.5\GeV$. Fraction
ratios $R^*$ are tabulated for $E_{\rm cut}\!=\!0.3\GeV$, 
$E_{\rm cut}\!=\!0.6\GeV$ and  $E_{\rm cut}\!=\!0.9\GeV$.

\subsection{Lepton energy moments}

\vspace*{4mm}

\hspace*{.2em}Table 1.~~{\small First moment of the lepton energy
$\aver{E_\ell}$.}
\vspace*{1mm}\\
\begin{tabular}{|c|l|l|l|l|l|l|l|l|}\hline
$ E_{\rm cut}  $& ~\hfill $ V $\hfill~ & \hfill~ $B$ \hfill~ & ~\hfill
$ C $ \hfill~
& ~\hfill $P$ \hfill~&~\hfill $D$ \hfill~&~\hfill $G$ 
\hfill~&\hfill~ $L $ \hfill~ & ~\hfill $S$ \hfill~ \\ \hline
$ 0 $&$1.372 $&$  0.389 $&$  -0.286 $&$  0.033  $&$ -0.085 $&$
-0.078 $&$  \msp{1.9}0.0043 $&$ -0.029
$ \\ \hline
$ 0.6 $&$1.420 $&$  0.367 $&$  -0.268 $&$  0.036 $&$  -0.083 $&$  -0.076
$&$ \msp{1.9}0.0035 $&$  -0.032
$\\ \hline
$ 0.9 $&$1.500$&$   0.339$&$   -0.243$&$   0.041$&$   -0.081$&$
   -0.073$&$ \msp{1.9}0.0021$&$   -0.035
$\\ \hline
$ 1.2 $&$1.614$&$   0.304$&$   -0.211$&$   0.051$&$   -0.083$&$
   -0.068$&$   -0.0005$&$   -0.043   $\\ \hline
$ 1.5 $&$ 1.759$&$  0.262$&$   -0.173$&$   0.073$&$   -0.096$&$
   -0.063$&$   -0.0058$&$   -0.055  $\\ \hline
\end{tabular}
\newpage

\noindent
\hspace*{.2em}Table 2.~~{\small Second moment of the lepton energy 
with respect to average, 
$\aver{(E_\ell\!-\!\aver{E_\ell})^2}$}\vspace*{1mm}\\
\begin{tabular}{|c|l|l|l|l|l|l|l|l|}\hline
$ E_{\rm cut}  $& ~\hfill $ V $\hfill~ & \hfill~ $B$ \hfill~ & ~\hfill
$ C $ \hfill~
& ~\hfill $P$ \hfill~&~\hfill $D$ \hfill~&~\hfill $G$ 
\hfill~&\hfill~ $L $ \hfill~ & ~\hfill $S$ \hfill~ \\ \hline
$ 0 $&$ 0.1774  $&$   0.0951 $&$    -0.0584 $&$    0.044 $&$ -0.059 
$&$ -0.029 $&$    -0.0048 $&$    -0.036 $\\ \hline
$ 0.6 $&$ 0.1391 $&$  0.0960$&$   -0.0602$&$   0.042$&$   -0.058$&$
   -0.027$&$   -0.0047$&$   -0.033  $\\ \hline
$ 0.9 $&$ 0.0969 $&$  0.0868$&$   -0.0547$&$   0.040$&$   -0.057$&$
   -0.024$&$   -0.0048$&$   -0.031 $\\ \hline
$ 1.2 $&$ 0.0570$&$   0.0691$&$   -0.0433$&$   0.038$&$   -0.055$&$
   -0.019$&$   -0.0051$&$   -0.029  $\\ \hline
$ 1.5 $&$ 0.0266$&$   0.0473$&$   -0.0293$&$   0.035$&$   -0.057$&$
   -0.012$&$   -0.0056$&$   -0.028   $\\ \hline
\end{tabular}\\
\vspace*{7mm}

\noindent
\hspace*{.2em}Table 3.~~{\small Third moment of the lepton energy 
with respect to average, 
$\aver{(E_\ell\!-\!\aver{E_\ell})^3}$}\vspace*{1mm}\\
\begin{tabular}{|c|l|l|l|l|l|l|l|l|}\hline
$ E_{\rm cut}  $& ~\hfill $ V $\hfill~ & \hfill~ $B$ \hfill~ & ~\hfill
$ C $ \hfill~
& ~\hfill $P$ \hfill~&~\hfill $D$ \hfill~&~\hfill $G$ 
\hfill~&\hfill~ $L $ \hfill~ & ~\hfill \hspace*{-3.7mm}
$100 \!\cdot\! S$ \hspace*{-3.2mm}\hfill~ \\ \hline
$ 0 $&$-0.0334 $&$  -0.0307 $&$  \msp{1.9}0.0265 $&$  0.023 $&$  -0.034 $&$
\msp{1.9}0.0043 $&$  -0.0054 $&$  -1.6 $\\ \hline
$ 0.6 $&$ -0.0121 $&$  -0.0204$&$   \msp{1.9}0.0189$&$   0.022$&$   -0.034$&$
 \msp{1.9}0.0021$&$ -0.0049$&$   -1.7 $\\ \hline
$ 0.9 $&$ -0.0015$&$   -0.0082$&$   \msp{1.9}0.0086$&$   0.019$&$
-0.033$&$ \msp{1.9}0.0003$&$   -0.0041$&$  -1.5
 $\\ \hline
$ 1.2 $&$\msp{1.9}0.0016$&$ \msp{1.9}0.0001$&$ \msp{1.9}0.0015$&$   0.015$&$
-0.030$&$   -0.0005$&$   -0.0031 $&$  -1.2 $\\ \hline
$ 1.5 $&$\msp{1.9}0.0009$&$   \msp{1.9}0.0030$&$   -0.0013$&$   0.010$&$
-0.025$&$   -0.0003\!$&$   -0.0020$&$   -0.9   $\\ \hline
\end{tabular}
\vspace*{7mm}

\noindent
\hspace*{.2em}Table 4.~~{\small Fraction of the decay rate $R^*$ with 
$E_\ell$ exceeding a threshold value $E_{\rm cut}$}\vspace*{1mm}\\
\begin{tabular}{|c|l|l|l|l|l|l|l|l|}\hline
$ E_{\rm cut}  $& ~\hfill $ V $\hfill~ & \hfill~ $B$ \hfill~ & ~\hfill
$ C $ \hfill~
& ~\hfill $P$ \hfill~&~\hfill $D$ \hfill~&~\hfill $G$ 
\hfill~&\hfill~ $L $ \hfill~ & ~\hfill\hspace*{-3.5mm}
$100 \!\cdot\! S$ \hspace*{-3mm}\hfill~ \\ \hline
$ 0.3 $&$0.9934$&$ 0.006$&$ -0.0044$&$ -0.0001$&$ -0.0008$&$
 -0.0008$&$ 0.0001$&$ 0.06 $\\ \hline
$ 0.6 $&$0.9508$&$ 0.040$&$ -0.0317$&$ -0.0008$&$ -0.0060$&$
 -0.0060$&$ 0.0009$&$ 0.15 $\\ \hline
$ 0.9 $&$0.8476$&$ 0.120$&$ -0.0947$&$ -0.0018$&$ -0.0176$&$
-0.0189$&$ 0.0029$&$ 0.19
 $\\ \hline
\end{tabular}
\vspace*{6mm}

\subsection{Hadron invariant mass moments}

Since the moments related to hadronic invariant mass and hadronic
energy are presently calculated using only 
$O(\alpha_s)$ perturbative corrections, we have employed
$\alpha_s=0.3$, which can equivalently be represented as 
\beq
\frac{\alpha_s(m_b)}{1-0.846\frac{9}{2} \frac{\alpha_s(m_b)}{\pi}}
\label{alpha}
\eeq
with the canonical value $\alpha_s(m_b)\!=\!0.22\;$. This allows
to use the same form Eq.~(\ref{310}) with $S$ showing the sensitivity to
perturbative corrections. In this way the interval $0.2\!<\!\alpha_s\!<\!0.45$
mentioned in Sect.~2 corresponds to varying $\alpha_s(m_b)$ within
$0.22^{+.07}_{-.06}$. Switching off perturbative corrections,
$\alpha_s\!=\!0$, numerically amounts to subtracting $0.16\,S$ from a moment. 

As seen from the tables, the dependence on the precise value of
$\alpha_s$ is moderate and the corresponding uncertainty is
subdominant; this is an advantage of using the Wilsonian scheme. 
\newpage

\noindent
\hspace*{.2em}Table 5.~~{\small First hadronic invariant mass moment 
$\aver{M_X^2}$.~~$S\!\simeq\! 0.41$}
\vspace*{1mm}\\
\begin{tabular}{|c|l|l|l|l|l|l|l|}\hline
$ E_{\rm cut}  $& ~\hfill $ V $\hfill~ & \hfill~ $B$ \hfill~ & ~\hfill
$ C $ \hfill~
& ~\hfill $P$ \hfill~&~\hfill $D$ \hfill~&~\hfill $G$ 
\hfill~&\hfill~ $L $ \hfill~  \\ \hline
$ 0 $&$4.641 $&$  -4.99$&$   3.18$&$   -0.70 $&$  1.0$&$   0.48$&$
   -0.13 $\\ \hline
$ 0.6 $&$4.619$&$   -4.96$&$   3.18$&$   -0.73$&$   1.0$&$
0.50$&$   -0.13 $\\ \hline
$ 0.9 $&$4.583$&$   -4.90$&$   3.19$&$   -0.79$&$   1.1$&$   0.55
$&$  -0.12 $\\ \hline
$ 1.2 $&$4.528$&$   -4.84$&$   3.21$&$   -0.93$&$   1.3$&$   
0.64$&$   -0.11 $\\ \hline
$ 1.5 $&$4.444$&$   -4.74$&$   3.22$&$   -1.29$&$   1.8$&$  0.81$&$
   -0.06 $\\ \hline
\end{tabular}
\vspace*{7mm}

\noindent
\hspace*{.2em}Table 6.~~{\small Second invariant mass moment
with respect to average, 
$\aver{(M_X^2\!-\!\aver{M_X^2})^2}$.~~$S\!\simeq\! -0.51$}
\vspace*{1mm}\\
\begin{tabular}{|c|l|l|l|l|l|l|l|}\hline
$ E_{\rm cut}  $& ~\hfill $ V $\hfill~ & \hfill~ $B$ \hfill~ & ~\hfill
$ C $ \hfill~
& ~\hfill $P$ \hfill~&~\hfill $D$ \hfill~&~\hfill $G$ 
\hfill~&\hfill~ $L $ \hfill~ \\ \hline
$ 0   $&$1.290$&$   0.396$&$   -0.97$&$  4.8$&$   -5.9$&$
-0.14$&$   \msp{1.9}0.30 $\\ \hline
$ 0.6 $&$1.233$&$   0.446$&$   -0.98 $&$  4.7$&$   -6.0$&$   -0.13
$&$  \msp{1.9}0.24  $\\ \hline
$ 0.9 $&$1.137$&$   0.523$&$   -0.99$&$   4.5$&$   -6.4$&$   -0.11
$&$   \msp{1.9}0.13
 $\\ \hline
$ 1.2 $&$0.985$&$   0.660$&$   -1.00$&$   4.2$&$   -7.1$&$   -0.06$&$
-0.05 $\\ \hline
$ 1.5 $&$0.747$&$   0.970$&$   -1.04$&$   3.8$&$   -8.9$&$
\msp{1.9}0.05$&$    -0.34 $\\ \hline
\end{tabular}
\vspace*{7mm}

\noindent
\hspace*{.2em}Table 7.~~{\small Third invariant mass moment
with respect to average, 
$\aver{(M_X^2\!-\!\aver{M_X^2})^3}$.~~$S\!\simeq\! 10.1$}
\vspace*{1mm}\\
\begin{tabular}{|c|l|l|l|l|l|l|l|}\hline
$ E_{\rm cut}  $& ~\hfill $ V $\hfill~ & \hfill~ $B$ \hfill~ & ~\hfill
$ C $ \hfill~
& ~\hfill $P$ \hfill~&~\hfill $D$ \hfill~&~\hfill $G$ 
\hfill~&\hfill~ $L $ \hfill~ 
\\ \hline
$ 0 $&$5.02$&$  1.40$&$   -2.50$&$   5.6$&$   21$&$   -1.3$&$
1.2 $\\ \hline
$ 0.6 $&$4.92$&$ 1.48$&$   -2.49$&$   5.5 $&$  21$&$
-1.3$&$    1.1  $\\ \hline
$ 0.9 $&$4.80$&$  1.53$&$   -2.45$&$   5.5$&$   19$&$   -1.3$&$
1.0 $\\ \hline
$ 1.2 $&$4.68$&$  1.52$&$   -2.32$&$   5.7$&$   17$&$   -1.2$&$
0.8  $\\ \hline
$ 1.5 $&$4.64$&$  1.40$&$   -1.98$&$   6.3$&$   15 $&$  -0.8 $&$
0.4  $\\ \hline
\end{tabular}
\vspace*{7mm}

\noindent
\hspace*{.2em}Table 8.~~{\small First modified hadronic moment 
$\aver{{\cal N}_X^2}$.~~$S\!\simeq\! 0.38$}
\vspace*{1mm}\\
\begin{tabular}{|c|l|l|l|l|l|l|l|}\hline
$ E_{\rm cut}  $& ~\hfill $ V $\hfill~ & \hfill~ $B$ \hfill~ & ~\hfill
$ C $ \hfill~
& ~\hfill $P$ \hfill~&~\hfill $D$ \hfill~&~\hfill $G$ 
\hfill~&\hfill~ $L $ \hfill~ 
\\ \hline
$ 0 $&$1.580$&$ -4.06$&$ 2.52$&$ -0.62$&$ 0.74$&$ 0.34$&$
 -0.10 $\\ \hline
$ 0.6 $&$1.579$&$ -4.03$&$ 2.53$&$ -0.64$&$ 0.77$&$ 0.36$&$
 -0.10 $\\ \hline
$ 0.9 $&$1.574$&$ -3.98$&$ 2.53$&$ -0.69$&$ 0.83$&$ 0.40$&$
 -0.09
 $\\ \hline
$ 1.2 $&$1.558$&$ -3.90$&$ 2.52$&$ -0.82$&$ 0.96$&$ 0.47$&$ -0.08  $\\ \hline
$ 1.5 $&$1.501$&$ -3.75$&$ 2.45$&$ -1.15$&$ 1.30$&$ 0.57$&$ -0.04  $\\ \hline
\end{tabular}
\newpage

\noindent
\hspace*{.2em}Table 9.~~{\small Second  modified hadronic moment with
respect to average 
$\aver{({\cal N}_X^2\!-\!\aver{{\cal N}_X^2})^2}$.~~$S\!\simeq\! -0.35$}
\vspace*{1mm}\\
\begin{tabular}{|c|l|l|l|l|l|l|l|}\hline
$ E_{\rm cut}  $& ~\hfill $ V $\hfill~ & \hfill~ $B$ \hfill~ & ~\hfill
$ C $ \hfill~
& ~\hfill $P$ \hfill~&~\hfill $D$ \hfill~&~\hfill $G$ 
\hfill~&\hfill~ $L $ \hfill~ 
\\ \hline
$ 0 $&$0.954$&$ 0.506$&$ -0.606$&$ 3.6$&$ -3.8$&$ 0.12$&$ 
\msp{1.9}0.25 $\\ \hline
$ 0.6 $&$0.918$&$ 0.537$&$ -0.616$&$ 3.5$&$ -3.9$&$ 0.12$&$ 
\msp{1.9}0.20  $\\ \hline
$ 0.9 $&$0.855$&$ 0.592$&$ -0.624$&$ 3.4$&$ -4.1$&$ 0.14$&$ 
\msp{1.9}0.13 $\\ \hline
$ 1.2 $&$0.761$&$ 0.685$&$ -0.628$&$ 3.2$&$ -4.6$&$ 0.18$&$
\msp{1.9}0.00  $\\ \hline
$ 1.5 $&$ 0.621$&$ 0.882$&$ -0.659$&$ 3.0$&$ -5.8$&$ 0.31$&$ -0.17 
$\\ \hline
\end{tabular}
\vspace*{7mm}

\noindent
\hspace*{.2em}Table 10.~~{\small Third  modified hadronic moment with
respect to average 
$\aver{({\cal N}_X^2\!-\!\aver{{\cal N}_X^2})^3}$.~~$S\!\simeq\! 9.9$}
\vspace*{1mm}\\
\begin{tabular}{|c|l|l|l|l|l|l|l|}\hline
$ E_{\rm cut}  $& ~\hfill $ V $\hfill~ & \hfill~ 
$\!\!\!\!B\!\!\!\!$ \hfill~ & ~\hfill
$ C \!\!\!$ \hfill~
& ~\hfill $P$ \hfill~&~\hfill $D$ \hfill~&~\hfill $G$ 
\hfill~&\hfill~ $L $ \hfill~ 
\\ \hline
$ 0 $&$3.27$&$ 1.47$&$ -1.67$&$ 2.5$&$ 13.3$&$ -0.87$&$ \msp{1.9}0.32 $\\ \hline
$ 0.6 $&$3.22$&$ 1.53$&$ -1.68$&$ 2.4$&$ 12.8$&$ -0.88$&$ \msp{1.9}0.29 $\\ \hline
$ 0.9 $&$3.15$&$ 1.57$&$ -1.66$&$ 2.5$&$ 12.1$&$ -0.89$&$ \msp{1.9}0.25
 $\\ \hline
$ 1.2 $&$3.10$&$ 1.57$&$ -1.58$&$ 2.7$&$ 11.1$&$ -0.88$&$ \msp{1.9}0.17  $\\ \hline
$ 1.5 $&$3.16$&$ 1.43$&$ -1.31$&$ 3.6$&$ \msp{.5}9.5$&$ -0.69$&$ -0.03  $\\ \hline
\end{tabular}
\vspace*{10mm}


Tables 11 and 12 give predictions for non-integer moments
$\aver{M_X}$ and  $\aver{M_X^3}$; they are evaluated using
Eq.~(\ref{180}) truncated after $k\!=\!3$.\vspace*{7mm}

\noindent
\hspace*{.2em}Table 11.~~{\small Hadronic invariant mass moment 
$\aver{M_X}$.~~$S\!\simeq\! 0.12$}
\vspace*{1mm}\\
\begin{tabular}{|c|l|l|l|l|l|l|l|l|}\hline
$ E_{\rm cut}  $& ~\hfill $ V $\hfill~ & \hfill~ $B$ \hfill~ & ~\hfill
$ C $ \hfill~
& ~\hfill $P$ \hfill~&~\hfill $D$ \hfill~&~\hfill $G$ 
\hfill~&\hfill~ $L $ \hfill~ 
\\ \hline
$ 0 $&$2.145$&$   -1.17$&$   0.750$&$   -0.22$&$   0.33$&$
   0.11$&$   -0.03 $\\ \hline
$ 0.6 $&$2.140$&$   -1.17$&$   0.752$&$   -0.22$&$   0.34$&$
0.12$&$   -0.03  $\\ \hline
$ 0.9 $&$2.133$&$   -1.16$&$   0.757$&$   -0.24$&$   0.37$&$
   0.13$&$   -0.03 
 $\\ \hline
$ 1.2 $&$2.122$&$   -1.15$&$   0.765$&$   -0.27$&$   0.42$&$
0.15$&$  -0.02  $\\ \hline
$ 1.5 $&$2.105$&$   -1.14$&$   0.770$&$   -0.35$&$   0.56$&$
0.19$&$ -0.01   $\\ \hline
\end{tabular}
\vspace*{7mm}

\noindent
\hspace*{.2em}Table 12.~~{\small Hadronic invariant mass moment 
$\aver{M_X^3}$.~~$S\!\simeq\! 1.1$}
\vspace*{1mm}\\
\begin{tabular}{|c|l|l|l|l|l|l|l|}\hline
$ E_{\rm cut}  $& ~\hfill $ V $\hfill~ & \hfill~ $B$ \hfill~ & ~\hfill
$ C $ \hfill~
& ~\hfill $P$ \hfill~&~\hfill $D$ \hfill~&~\hfill $G$ 
\hfill~&\hfill~ $L $ \hfill~ 
\\ \hline
$ 0 $&$10.19$&$ -16.0$&$ 10.1$&$ -1.4$&$ 2.1$&$ 1.5$&$ -0.38
$\\ \hline
$ 0.6 $&$10.11$&$ -15.8$&$ 10.1$&$ -1.5$&$ 2.1$&$ 1.6$&$ -0.38
$\\ \hline
$ 0.9 $&$\msp{.6}9.98$&$ -15.6$&$ 10.1$&$ -1.8$&$ 2.3$&$ 
1.7$&$ -0.38
$\\ \hline
$ 1.2 $&$\msp{.6}9.78$&$ -15.2$&$ 10.1$&$ -2.2$&$ 2.8$&$ 2.0$&$ -0.37
$\\ \hline
$ 1.5 $&$\msp{.6}9.47$&$ -14.8$&$ 10.0$&$ -3.4$&$ 3.9$&$ 
2.6$&$ -0.27
$\\ \hline
\end{tabular}
\vspace*{7mm}

Below we give the tables for the higher integer hadronic moments (for mass
squared and modified) with respect to a fixed mass. \vspace*{7mm}

\noindent
\hspace*{.2em}Table 13.~~{\small Second invariant mass moment
with respect to $4\GeV^2$, 
$\aver{(M_X^2\!-\!4\GeV^2)^2}\,$.\,\footnote{There are strong
cancellations in the perturbative corrections
here, and the face value of $S$ is not meaningful. One can roughly 
assume $|S| \!\lsim\! 0.25$.}
}
\vspace*{1mm}\\
\begin{tabular}{|c|l|l|l|l|l|l|l|l|}\hline
$ E_{\rm cut}  $& ~\hfill $ V $\hfill~ & \hfill~ $\!B\!\!$ 
\hfill~ & ~\hfill$ \!C \!\!\!$ \hfill~
& ~\hfill $P$ \hfill~&~\hfill $\!D\!\!$ \hfill~&~\hfill $G$ 
\hfill~&\hfill~ $L $ \hfill~ 
\\ \hline
$ 0 $&$ 1.701$&$ -5.76$&$ 3.20$&$ 3.9$&$ -4.6$&$ 0.47$&$ \msp{1.9}0.13 
  $\\ \hline
$ 0.6 $&$1.617$&$ -5.45$&$ 3.06$&$ 3.8$&$ -4.7$&$ 0.50$&$ \msp{1.9}0.08 
  $\\ \hline
$ 0.9 $&$1.477$&$ -4.95$&$ 2.83$&$ 3.6$&$ -5.0$&$ 0.54$&$ -0.02
  $\\ \hline
$ 1.2 $&$  1.264$&$ -4.21$&$ 2.49$&$ 3.2$&$ -5.7$&$ 0.62$&$ -0.17
  $\\ \hline
$ 1.5 $&$0.944$&$ -3.02$&$ 1.92$&$ 2.7$&$ -7.4$&$ 0.78$&$ -0.40
  $\\ \hline
\end{tabular}
\vspace*{7mm}

\noindent
\hspace*{.2em}Table 14.~~{\small Third invariant mass moment
with respect to $4\GeV^2$, 
$\aver{(M_X^2\!-\!4\GeV^2)^3}$.~~$S\!\simeq\! 11$}
\vspace*{1mm}\\
\begin{tabular}{|c|l|l|l|l|l|l|l|l|}\hline
$ E_{\rm cut}  $& ~\hfill $ V $\hfill~ & \hfill~ $B$ \hfill~ & ~\hfill
$ C $ \hfill~
& ~\hfill $P$ \hfill~&~\hfill $D$ \hfill~&~\hfill $G$ 
\hfill~&\hfill~ $L $ \hfill~ 
\\ \hline
$ 0 $&$7.77$&$ -22.9$&$ 12.0$&$ 11.2$&$ 15.0$&$ 0.85$&$ \msp{1.9}1.09   
  $\\ \hline
$ 0.6 $&$ 7.45$&$ -21.3$&$ 11.2$&$ 10.6$&$ 14.2$&$ 0.88$&$ \msp{1.9}0.91
  $\\ \hline
$ 0.9 $&$6.98$&$ -18.9$&$ 10.0$&$ \msp{.5}9.7$&$ 12.8$&$ 0.96$&$ \msp{1.9}0.63
  $\\ \hline
$ 1.2 $&$ 6.39$&$ -15.5$&$ \msp{.5}8.3$&$ \msp{.5}8.7$&$ 10.8$&$ 
1.15$&$ \msp{1.9}0.25
  $\\ \hline
$ 1.5 $&$5.72$&$ -10.6$&$ \msp{.5}5.8$&$ \msp{.5}7.6$&$ \msp{.5}7.5$&$ 
1.62$&$ -0.27  
  $\\ \hline
\end{tabular}
\vspace*{7mm}

\noindent
\hspace*{.2em}Table 15.~~{\small Second modified 
moment with respect to $1.35\GeV^2\!$, 
$\aver{({\cal N}_X^2\!-\!1.35\GeV^2)^2}$.~$\;S\!\simeq\! 
-0.19$}\hspace*{-.5mm} \vspace*{1mm}\\
\begin{tabular}{|c|l|l|l|l|l|l|l|l|}\hline
$ E_{\rm cut}  $& ~\hfill $ V $\hfill~ & \hfill~ $B$ \hfill~ & ~\hfill
$ C $ \hfill~
& ~\hfill $P$ \hfill~&~\hfill $D\!$ \hfill~&~\hfill $G$ 
\hfill~&\hfill~ $L $ \hfill~ 
\\ \hline
$ 0 $&$1.007$&$ -1.20$&$ 0.62$&$ 3.3$&$ -3.4$&$ 0.28$&$ \msp{1.9}0.21   
  $\\ \hline
$ 0.6 $&$ 0.970$&$ -1.14$&$ 0.60$&$ 3.2$&$ -3.5$&$ 0.29$&$ \msp{1.9}0.16
  $\\ \hline
$ 0.9 $&$0.905$&$ -1.03$&$ 0.57$&$ 3.1$&$ -3.7$&$ 0.32$&$ \msp{1.9}0.08
  $\\ \hline
$ 1.2 $&$ 0.805$&$ -0.78$&$ 0.49$&$ 2.9$&$ -4.2$&$ 0.38$&$ -0.03
  $\\ \hline
$ 1.5 $&$0.644$&$ -0.11$&$ 0.14$&$ 2.7$&$ -5.4$&$ 0.49$&$ -0.19
  $\\ \hline
\end{tabular}
\vspace*{7mm}

\noindent
\hspace*{.2em}Table 16.~~{\small Third modified 
moment with respect to $1.35\GeV^2$, 
$\aver{({\cal N}_X^2\!-\!1.35\GeV^2)^3}$.~~$S\!\simeq\! 11$}
\begin{tabular}{|c|l|l|l|l|l|l|l|l|}\hline
$ E_{\rm cut}  $& ~\hfill $ V $\hfill~ & \hfill~ $B$ \hfill~ & ~\hfill
$ C $ \hfill~
& ~\hfill $P$ \hfill~&~\hfill $D$ \hfill~&~\hfill $G$ 
\hfill~&\hfill~ $L $ \hfill~ 
\\ \hline
$ 0 $&$ 3.94$&$ -10.4$&$ 5.5$&$ 3.0$&$ 12.8$&$ 0.25$&$ \msp{1.9}0.19
  $\\ \hline
$ 0.6 $&$ 3.86$&$ -9.8$&$ 5.2$&$ 2.9$&$ 12.3$&$ 0.26$&$ \msp{1.9}0.15 
  $\\ \hline
$ 0.9 $&$3.74$&$ -8.8$&$ 4.8$&$ 2.8$&$ 11.5$&$ 0.28$&$ \msp{1.9}0.08
  $\\ \hline
$ 1.2 $&$3.59$&$ -7.4$&$ 4.1$&$ 2.7$&$ 10.4$&$ 0.36$&$ -0.02
  $\\ \hline
$ 1.5 $&$3.45$&$ -5.5$&$ 3.1$&$ 2.6$&$  \msp{.5}9.1$&$ 0.56$&$ -0.18
  $\\ \hline
\end{tabular}

\end{document}